\documentclass[aps,twocolumn,superscriptaddress,floatfix,letter]{revtex4}

\usepackage{graphicx}
\usepackage{amsmath}
\usepackage{hyperref}
\usepackage{url}
\usepackage{float}
\usepackage[caption = false]{subfig}

\newcommand{\widebar}[1]{\mkern 1.5mu\overline{\mkern-1.5mu#1\mkern-1.5mu}\mkern 1.5mu}
\newcommand{\ket}[1]{\left\vert #1\,\right\rangle}
\newcommand{\bra}[1]{\left\langle #1\,\right\vert}

\begin{document}

\setlength\abovedisplayskip{5pt}
\setlength\belowdisplayskip{5pt}
\title{Temperature of a single chaotic eigenstate}

\author{Fausto Borgonovi}
\affiliation{Dipartimento di Matematica e
  Fisica and Interdisciplinary Laboratories for Advanced Materials Physics,
  Universit\`a Cattolica, via Musei 41, 25121 Brescia, Italy}
\affiliation{Istituto Nazionale di Fisica Nucleare,  Sezione di Pavia,
  via Bassi 6, I-27100,  Pavia, Italy}
\author{Francesco Mattiotti}
\affiliation{Dipartimento di Matematica e
  Fisica and Interdisciplinary Laboratories for Advanced Materials Physics,
  Universit\`a Cattolica, via Musei 41, 25121 Brescia, Italy}
\author{Felix M. Izrailev}
\affiliation{Instituto de F\'{i}sica, Benem\'{e}rita Universidad Aut\'{o}noma
  de Puebla, Apartado Postal J-48, Puebla 72570, Mexico}
\affiliation{NSCL and Dept. of Physics and Astronomy, Michigan State University, E. Lansing, Michigan 48824-1321, USA}

\date{\today}

\begin{abstract}
The onset of thermalization in a closed  system of randomly interacting bosons, at the level of a single eigenstate, is discussed. We focus on the emergence of  Bose-Einstein distribution of single-particle occupation numbers, and we give a local criterion for thermalization dependent on the  eigenstate energy. We show how to define the temperature of an eigenstate, provided that it has a chaotic structure in the basis defined by the single-particle states. The analytical expression for the eigenstate temperature as a function of both  inter-particle interaction and energy is complemented by numerical data.
\end{abstract}

\maketitle

\section{Introduction}

The subject of thermalization occurring in isolated quantum systems of interacting particles has been developed the last decades due to various applications in nuclear and atomic physics~\cite{zele,flam}, as well as in view of basic problems of statistical mechanics~\cite{rep,basic,basic1}. Recently the interest to this subject has increased due to experiments with cold atoms and molecules in optical lattices~\cite{exp} and  trapped ions~\cite{exp1}. Correspondingly, many theoretical and numerical studies have been performed in order to understand the mechanism of thermalization in the absence of a heat bath (see ~\cite{rep} and Refs therein). Nevertheless, despite some studies about the onset of thermalization as a function of various physical parameters such as the number of particles\cite{manan}, the strength of inter-particle interaction\cite{our2012}, the choice of initially excited states~\cite{leajon},  the role of these items still remains open.

The mechanism driving thermalization in isolated systems of interacting particles is associated with quantum chaos~\cite{our2016}. Differently from the well developed one-body chaos theory, many problems related to many-body chaos, such as the thermalization of Fermi and Bose particles, are currently under intensive studies. Unlike classical chaos which is intrinsically related to the instability of motion with respect to a change of initial conditions, quantum chaos manifests itself  in specific fluctuations of  the energy spectra and in the chaotic structure of  eigenstates. As shown in \cite{our2012}, the properties of the energy spectra are less important to the statistical relaxation toward a steady-state distribution than the structure of the eigenstates in the physically chosen many-particle basis. Therefore, the main interest in the study of many-body chaos was shifted, since long, to the properties of many-body eigenstates.

Chaotic eigenstates play a key role in the statistical description of isolated quantum systems.  As  stressed long ago \cite{Landau}, conventional statistical mechanics can be established on the level of individual quantum states and not only by averaging over many states. This has been confirmed numerically decades ago, see for instance \cite{our2016}  and discussion in \cite{jen}.
However, this fact has no practical consequences unless the conditions for such a situation are developed. One of the open problems in this field is to establish these conditions for systems with a finite number of particles.

To date, many problems have been addressed concerning the problem  of thermalization in isolated systems. Here we raise a new one, which is directly related to this issue. It is already agreed that one can speak of thermalization on the level of an individual state, and various characteristics of  thermalization have been under extensive studies, such as the relaxation of a system to steady state distributions after various quenches, decay of correlations in time for observables and their fluctuations after  relaxation, etc.~\cite{exp,exp1,BBIS04,manan,leajon,our2012}.

Now, in view of the basic concepts of statistical mechanics and recent experiments with cold atoms and molecules\cite{exp}
 it is natural to ask the question about the onset of the Bose-Einstein distribution (BED) emerging on the level of a {\it single} many-body eigenstate,  due to the interaction between bosons and not to an external field or a  thermostat.  Below we specifically initiate the study of the onset of BED in a 
finite system  of interacting bosons, that is expected to occur when the inter-particle interaction is strong enough. We suggest a semi-analytical approach able to reveal the conditions under which 
an isolated many-body eigenstate can be considered thermal and 
  introduce its temperature  in relation to the model parameters.

\section{The model and basic concepts }

At variance with eigenvalues, many-particle eigenstates are defined by means of a suitable single-particle basis. The latter, from its side has a direct relevance to the physical reality, specifically, to the choice of the mean field to which quantum observables such as occupation numbers are referred to. Correspondingly, we assume that the total Hamiltonian $H$ can be presented as the sum of the mean field  $H_0$ describing non-interacting (quasi) particles, and a residual interaction $V$, modeled  as a two-body random interaction. Such a setup, based on a random interaction, also serves as a good model for a deterministic interaction between bosons\cite{BBIS04} where the complexity in many-body matrix elements emerges due to the  complicated nature of the interaction itself.

In this paper
we consider $N$ identical bosons occupying $M$ single-particle levels specified by random energies $\epsilon_s$ with mean spacing, $\langle \epsilon_s- \epsilon_{s-1} \rangle = 1 $. 
Let us notice that the randomness in the single particle spectrum is not strictly necessary
for the results obtained: it has been introduced only in order to avoid the degeneracies in the
unperturbed many-body spectrum.

The Hamiltonian reads
\begin{equation}
  H= H_0 + V = \sum_s \epsilon_s \, a^\dag_s a_s  +
 \sum_{s_1 s_2 s_3 s_4} V_{s_1 s_2 s_3 s_4} \, a^\dag_{s_1} a^\dag_{s_2} a_{s_3} a_{s_4}
\label{ham}
\end{equation}
where the two-body matrix elements $ V_{s_1 s_2 s_3 s_4} $ are random Gaussian entries with  zero mean and variance $V^2$. The dimension of the Hilbert space generated by the many-particle basis states is ${N_H} = (N+M-1)!/N!(M-1)!$. Here we consider  $N=6$ particles in $M=11$ levels (dilute limit, $N \leq M$) for which $N_H = 8008$. 

Two-body random interaction (TBRI) matrices (\ref{ham}) have a quite long history. They were introduced  in \cite{TBRI}  and extensively studied for fermions \cite{fermi}. On the other hand the case of Bose particles has been less investigated and 
 only few results are known and typically for the dense limit,  $N \gg M$ \cite{bosons,kota-bose}.

The eigenstates of $H$ can be generically 
represented in terms of the basis states $|k\rangle = a^\dagger_{k_1}...a^\dagger_{k_N} |0\rangle$ 
which are eigenstates of $H_0$,
\begin{equation}
\ket{\alpha} = \sum_k C_k^{\alpha} \ket{k} ,
\label{ldos}
\end{equation}
where it has been implicitly assumed that
\begin{equation}
 H|\alpha \rangle = E^\alpha |\alpha \rangle,  
\label{ldos1}
\end{equation}
and 
\begin{equation}
  H_0 |k\rangle = E^0_{k} |k\rangle .
\label{ldos2}
\end{equation}
A characterization of the number of principal components  $C_k^{\alpha}$ in an eigenstate $\ket{\alpha}$
can be obtained by the study of the participation ratio,
\begin{equation}
N_{pc}=1/\sum_k |C_k^\alpha|^4.
\label{ipr}
\end{equation}
If the number $N_{pc}$ of the principal components $C_k^{\alpha}$  is sufficiently large (we will specify later how much large it should be) and $C_k^{\alpha}$ can be considered as random and non-correlated ones, this is the case of {\it chaotic} eigenstates. This notion is quite different from full delocalization in the unperturbed basis since for isolated systems the eigenstates typically fill only a part of the unperturbed basis \cite{our2016}.

 To characterize the structure of the eigenstates, we use the {\it $F-$function},
\begin{equation}
F^\alpha (E) = \sum_k |C_k^\alpha |^2 \delta(E-E^0_k),
\end{equation}
which is the energy representation of an eigenstate. From the components $C_k^{\alpha}$ one can  also construct  the {\it strength function} (SF) of a basis state $\ket{k}$,
\begin{equation}
F_k(E) = \sum_{\alpha} |C^{\alpha}_k|^2 \delta (E-E^{\alpha})
\label{ld1}
\end{equation}
widely used in nuclear physics \cite{SF} and known in solid state physics as {\it local density of states}. The SF shows how the basis state $\ket{k}$ decomposes into the exact eigenstates $\ket{\alpha}$ due to the interaction $V$. It can be measured experimentally and it is of great importance since its Fourier transform gives the time evolution of an excitation initially concentrated in the basis state $\ket{k}$. Specifically, it defines the survival probability to find the system at time $t$ in the initial state $\ket{k}$.

On increasing the interaction strength, the SF in isolated systems undergoes a crossover from a delta-like function (perturbative regime) to a Breit-Wigner (BW), with a width well described by the Fermi golden rule. With a further increase of the interaction, the form of the SF tends to a Gaussian\cite{basic,our2012,ljnjp}, a scenario that has also been  observed   experimentally~\cite{horacio}.

One of the basic concepts in our approach is the so-called {\it energy shell} which is the energy region defined by the projection of $V$ onto the basis of $H_0$ \cite{chirikov}. This region is the largest one that can be occupied by an eigenstate. The partial filling of the energy shell by an eigenstate can be associated with the many-body localization in the energy representation, a subject that is nowadays  under intensive investigation (see, for example, \cite{MBL} and references therein). When this happens, of course, the eigenstates cannot be treated as thermal, in the sense that a good definition of temperature cannot be done. Contrary to this, if a chaotic eigenstate fills completely the energy shell, this corresponds to maximal quantum chaos, and  a proper temperature can be defined.

In the past a  parameter driving the global crossover from non-chaotic to chaotic eigenstates has been proposed based essentially on the ratio between the interaction strength and the mean energy range spanned by the basis states effectively coupled by the interaction $V$\cite{basic, basic1, our2016}. This criterion is independent of the energy of the eigenstate.  Since we are dealing here with single eigenstates we will generalize this idea in order to obtain a local criterion (i.e. depending on the eigenenergy) for such a crossover.

Each many--body eigenstate $ \ket{\alpha}$ is not only characterized by an ``effective number'' of occupied basis states,
$ N_{pc}$ i.e. a number of principal components in the unperturbed basis,  but also by 
an  unperturbed energy width,    
\begin{equation}
\label{delta0}
\delta_0 = \left(\bra{\alpha} H_0^2 \ket{\alpha} - \bra{\alpha} H_0 \ket{\alpha}^2\right)^{1/2}.
\end{equation}
 These two parameters allow to define, for each single eigenstate,  an effective mean energy spacing, $ d_{loc} = \delta_0/N_{pc} $, that the perturbation strength $V$ must overcome in order to go beyond the perturbative regime. Accordingly, in order to have chaotic eigenstates we require  $V > d_{loc}$, while for $V < d_{loc}$ we can speak of perturbative regime. In the following we will see that the region
characterized by $V > d_{loc}$ is the ``thermal '' one,  where an effective  temperature,
dependent on the inter-particle interaction,
can be defined via the Bose-Einstein distribution.

\section{The Bose-Einstein distribution for an interacting eigenstate} 

In order to define the temperature for each selected eigenstate $|\alpha\rangle$ let us consider its occupation number distribution (OND),
\begin{equation}
\label{spond}
n_s^\alpha = \langle \alpha | \hat{n}_s | \alpha \rangle = \sum_k |C_k^\alpha|^2
\langle k | \hat{n}_s | k \rangle.
\end{equation}
As one can see, the OND (\ref{spond}) consists of two ingredients: the probabilities $|C_k^\alpha|^2$ and the occupation numbers $\langle k | \hat{n}_s | k \rangle$ related to the basis states of $H_0$. The latter are just integer numbers $0,1,2,...N$ depending on how many bosons occupy the single-particle level $s$ with respect to the many-body state $| k \rangle $. If the eigenstate $| \alpha \rangle$ of $H$ consists of many uncorrelated components one can substitute $|C_k^\alpha|^2$ by the corresponding SF obtained either by an average over a number of eigenstates with close energies, or {\it inside an individual eigenstate}, for example, with the use of the ``moving window" average \cite{our2016}. Thus,  from the knowledge of the SF it is possible to obtain the OND without the diagonalization of  huge Hamiltonian matrices.

\begin{figure}[t]
\includegraphics[width=7cm]{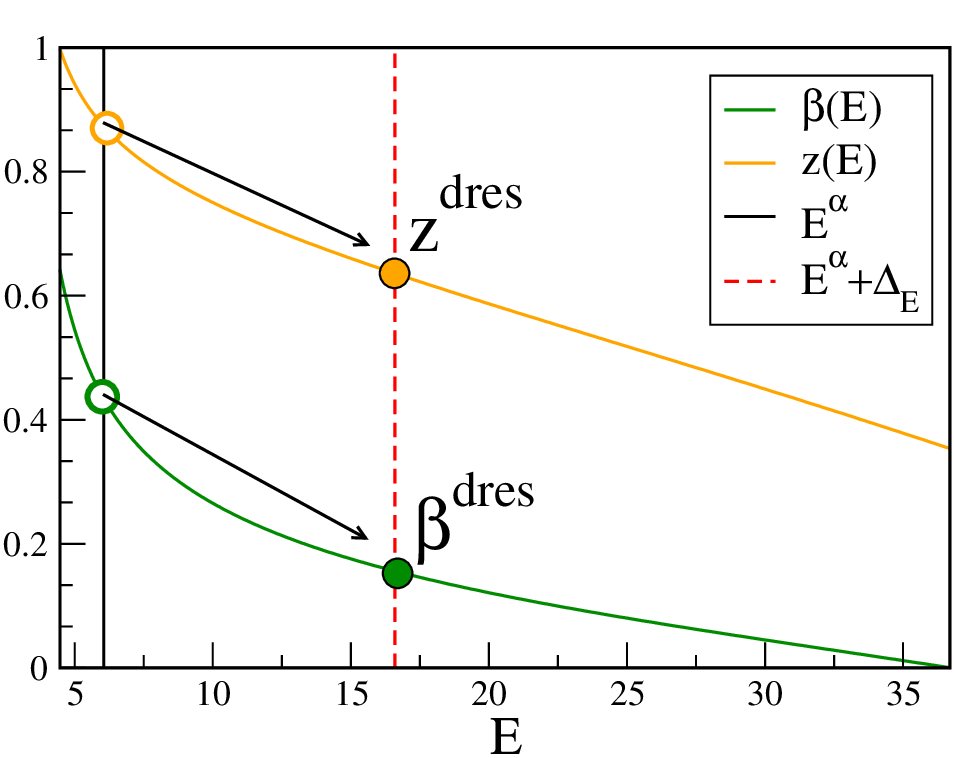}
  \caption{(Color online) Pictorial description of the increase of temperature for the 
eigenstate d) in Fig.~\ref{fmi04}.
 }
\label{fmi03b}
\end{figure}

\begin{figure}[t]
\includegraphics[width=8cm]{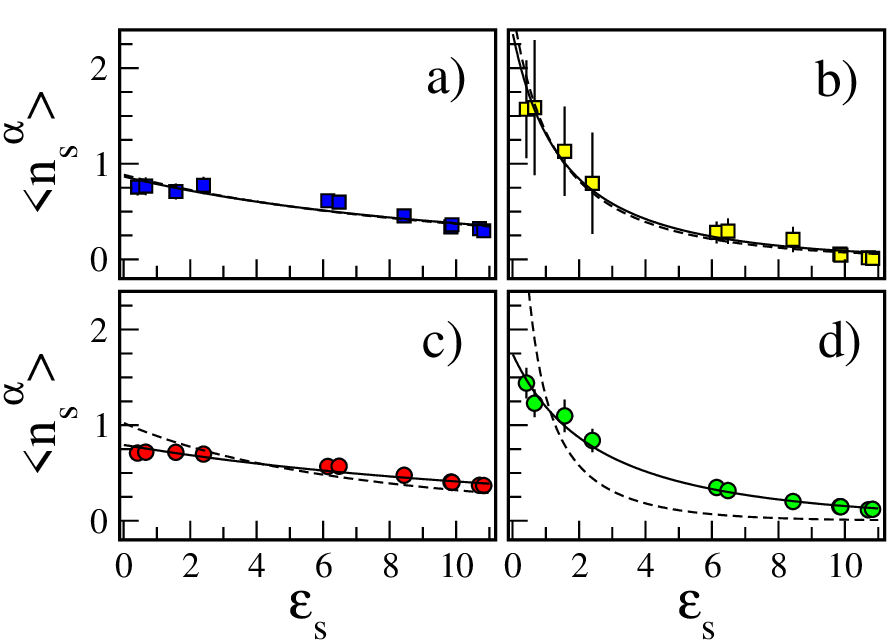}
\caption{(Color online) Average occupation numbers $\langle n_s^\alpha \rangle $ for weak ($V=0.1$, top panels) and strong ($V=0.4$, bottom panels) perturbation and different energies:  middle of  the energy spectrum  : a) $E^\alpha = 28.51$,  c) $E^\alpha = 25.93$; edges : b) $E^\alpha = 11.27$ and  d) $E^\alpha = 6.05$. Error bars indicate one standard deviation and are obtained by averaging over 20 close eigenstates. Dashed curves are obtained from the BED with $E=E^\alpha$ in Eq.~(\ref{fco}). Full curves correspond to the BED with the energy $E^{dres}$ in Eq.~(\ref{she}).}
\label{fmi04}
\end{figure}

Having in mind to define the temperature of a single eigenstate by means of its corresponding OND, few relevant questions come out. First of all, since we are dealing with bosons, the 
common reference OND is the Bose-Einstein distribution (BED) that is derived for non-interacting particles in the thermodynamic limit. The situation here is clearly different since our system has a finite number of interacting particles. To address properly this question we start with the basic relations,
\begin{equation}
\sum_s n_s = N
\qquad \sum_s \epsilon_s n_s = E ,
\label{fco}
\end{equation}
where $N$ is the total number of bosons and $E$ is the energy of a system for which  the inter-particle interaction is neglected. As is known, the solution of these equations for $N \rightarrow \infty $ leads to the BED,
\begin{equation}
n_s^{BE}= (e^{\beta(\epsilon_s-\mu)}-1)^{-1}.
\label{fbe}
\end{equation}
The derivation can be obtained due to the combinatorics only, with the constants $\beta$ and $\mu$ as the Lagrange multipliers \cite{RR80} (see, also, discussion in Ref.\cite{TYW16}). The meaning of $\beta$ and $\mu$ as the inverse temperature and chemical potential respectively, emerges when the system is connected with a heat bath. However, we will show that one can speak of BED even if the system is isolated; moreover, this distribution emerges on the level of a single eigenstate of the total Hamiltonian. Inserting (\ref{fbe}) into (\ref{fco}), one can  obtain both $\beta$ and $z=e^{\beta\mu}$ as a function of  $N$ and $E$. 
  If we further fix the number
of particles $N$ we obtain two functions $z(E) $ and $\beta(E)$, as shown
in Fig.~\ref{fmi03b}. The  values of $z$ and $\beta$ correspondent to 
the energy $E^\alpha$ are indicated in Fig.~\ref{fmi03b} 
by empty circles that are obtained by the intersection of the vertical line $E=E^\alpha$
with  the curves $z(E)$ and $\beta(E)$. 
Let us note that the BED
 indicated by a dashed line in Fig.~\ref{fmi04} d)
has been obtained using exactly these values  of $z$ and $\beta$.

Now the key question is: what  energy in the r.h.s. of Eq.~(\ref{fco}) should be used for  interacting bosons in order to have, if any, the correspondence to the numerically obtained OND \cite{fausto} ?

First, we start with the global correspondence between the actual OND numerically obtained from individual eigenstates (\ref{spond}) and the BED 
expression (\ref{fbe}). For this we consider the OND averaged over a number of close eigenstates in a narrow energy window. 
We considered the average over a small energy window with the only purpose to study fluctuations in the next section. In any case
the OND's for  single eigenstates are shown in Fig.\ref{fmi04single} in Appendix.

We choose the eigenenergy $E^\alpha$ in two different regions: close to the center and to the edges of the spectrum, and, for each of them, two different values of the interaction strength $V$, see Fig.~\ref{fmi04}. 
In each panel of this figure   the  average values for the OND are shown  with the error bars
representing one standard deviation (fluctuations here are due to different eigentates in a close energy window, 
alternatively one can choose one single eigenstate and change the random inter-particle potential) and two curves,  dashed and full.
The dashed curves are those obtained by choosing as $E_\alpha$ the unperturbed energy while the full ones are obtained by ``dressing''
the energy, as shown here below.

As one can see in Fig.~\ref{fmi04},
while for weak interaction (top panels) dashed lines  match perfectly numerical data,  this does not happen for strong interaction (bottom panels).  While such a failure in case of strong interaction is not unexpected, the good agreement in the case of weak interaction is far from trivial, since, it is worth noting that the  Bose-Einstein
distribution  is obtained in the limit $N\to \infty$ while here we have only $N=6$ particles!

To take into account the inter-particle interaction we use the approach suggested in Refs~\cite{basic,basic1}. Specifically, we substitute the energy $E=E^{\alpha}$ in (\ref{fco}) with the "dressed" energy,
\begin{equation}
E^{dres} = \langle \alpha | H_0 | \alpha \rangle \equiv  E^{\alpha} + \Delta_\alpha .
\label{she}
\end{equation}
Note that this energy is higher (in the region in which the density of states (DOS)  increases with energy) than the eigenvalue $E^\alpha$ corresponding to the eigenstate $\ket{\alpha}$. This corresponds to a temperature $T^{dres}$ higher than that obtained with the substitution $E \to E^\alpha$. The dressed energy $E^{dres} = E^\alpha + \Delta_\alpha$  has been 
indicated as a vertical dashed  line in  Fig.~\ref{fmi03b}. Since 
the energy shift $\Delta_\alpha$  is always positive in the energy region
where the  density of states increases with  the energy,
this produces a lowering of 
both $z$ and $\beta$ indicated in Fig.~\ref{fmi03b} 
as full circles ($z^{dres}$ and $\beta^{dres}$).

 Plugging the BED , Eq.~(\ref{fbe}), in  Eq.~(\ref{fco}) with the substitution  $E^{dres} \to E $  returns the values of  $\mu$ and $\beta$ 
from which we can write down the  corresponding BED indicated by full curves
  in Fig.~\ref{fmi04}. Even if, in the case of weak interaction (top column), the BED is hardly distinguishable from the ``unperturbed one'', for strong interaction (bottom panels)  they 
are very different, nevertheless, they match the numerical data extremely well, without any fit!

\begin{figure}[t]
\includegraphics[width=8cm]{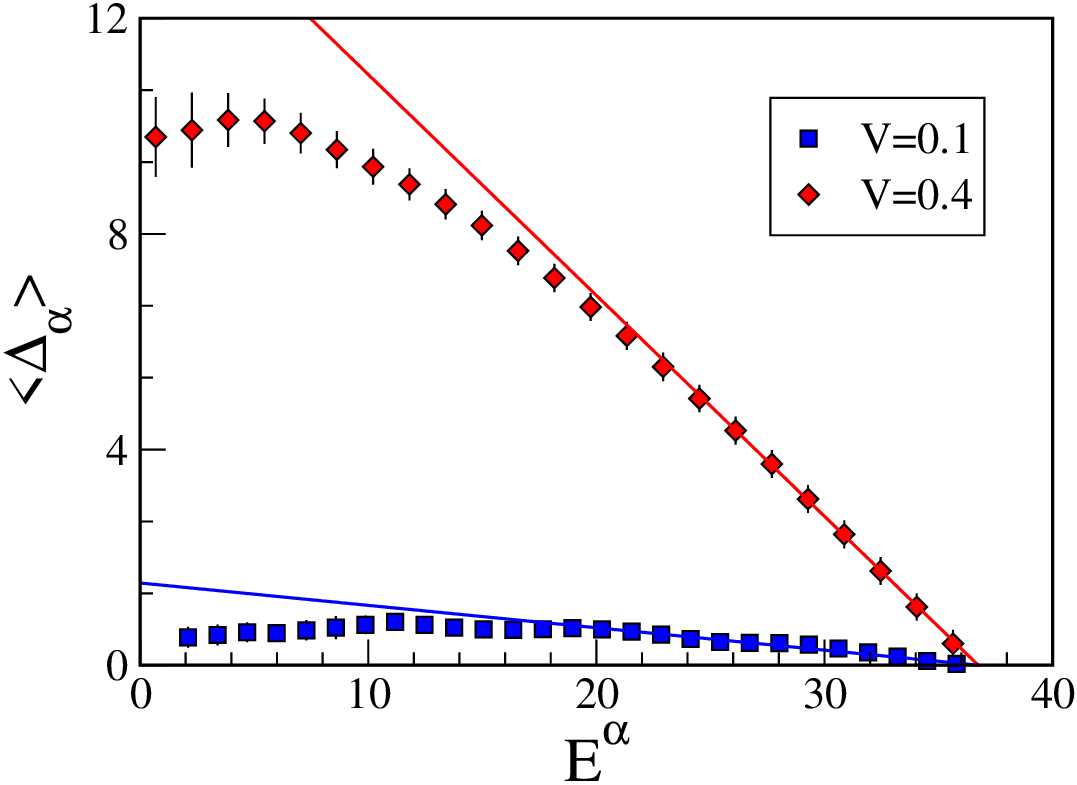}
  \caption{ (Color online)
Average energy shift $\langle \Delta_\alpha \rangle $   as a function of the energy $E^\alpha$
for two different values of the interaction $V$.
Symbols stand for numerical results while full lines represent
the Gaussian approximation.
(error bars indicate one standard deviation).
Average has been done over $20$ close eigenstates.
Due to the symmetry 
only half of the energy spectrum (where the density of states is an increasing function of the energy)
is shown.
 }
\label{fmi03}
\end{figure}

The energy shifts $\Delta_\alpha$ can be easily calculated numerically for each eigenstate.
In Fig.~\ref{fmi03} we plot such values, averaged over close eigenstates for
the two different perturbation strengths $V$ considered in Fig.~\ref{fmi04}. 
 
It is also possible to derive
an analytical expression for the energy shift $\Delta_\alpha$ in  Eq.~(\ref{she})  under not too strong assumptions. For weak TBRI and  a  large number of particles, the form of  the density of states (DOS)  is a  Gaussian \cite{gauss}. Moreover, due to the trace conservation of $H$, the position of the center $E_c$ of the perturbed spectrum is the same of the unperturbed one. In this situation  the variance $\sigma_E$ of the ``perturbed'' DOS $\rho$ is related to the variance $\sigma_0$  of the ``unperturbed'' DOS $\rho_0$ 
according to the simple relation (see Appendix  for details) 
\begin{equation}
\label{sig0}
\sigma_E^2 = \sigma_0^2 + \widebar{(\Delta E)}^2
\end{equation}
 where 
\begin{equation}
\label{deltae2}
\widebar{(\Delta E)^2} = (1/N_H) \sum_n \sum_{k\ne n} H^2_{nk}
\end{equation}
  is the average width of the SF and it can be obtained without any diagonalization. 
Inserting in   Eq.~(\ref{she}),  the spectral decomposition of $H_0$,
\begin{equation}
\label{sd}
H_0  = \sum_k E^0_k |k\rangle \langle k |
\end{equation}
one has,
\begin{equation}
\label{es1}
\Delta_\alpha = \sum_k E^0_k |C_k^\alpha|^2  - E^\alpha \simeq \int  dE (E-E^\alpha) \rho_0(E) \langle |C_k^\alpha|^2\rangle.
\end{equation}
Assuming  a Gaussian form also for  $\langle |C_k^\alpha|^2\rangle$ peaked around $E^\alpha$,
\begin{equation}
\label{gau1}
 \langle |C_k^\alpha|^2\rangle \simeq  \exp[-(E-E^\alpha)^2/2\widebar{(\Delta E)^2}]
\end{equation}
and for $\rho_0(E)$,
\begin{equation}
\label{gau2}
 \rho_0(E)  \simeq  \exp[-(E-E_c)^2/2\sigma_0^2]
\end{equation}
and inserting in Eq.~(\ref{es1}) with the correct normalizations, one gets 
the analytical estimate for the energy shift:
\begin{equation}
\label{es2}
\Delta_\alpha = 
\displaystyle \frac{\widebar{(\Delta E)}^2}{\widebar{(\Delta E)}^2 + \sigma_0^2}
\left( E_c - E^\alpha \right)
\end{equation}
These analytical values have been shown in Fig.~\ref{fmi03} as solid lines. As one can see
they  work very well in the center of the energy spectrum (where 
the hypothesis of Gaussian
LDOS and DOS can be applied without appreciable errors) while significant deviations appear at the low edge of the spectrum, where,
due to the finite number of particles and levels it is well  known that both DOS and LDOS cannot be 
described by Gaussians.

The increase of temperature, $\Delta T$, emerging due to the inter-particle interaction, can  be obtained from the definition of thermodynamic temperature,  by means of  the unperturbed density of states $\rho_0$,
$$\beta = \frac{1}{T}= \frac{d\ln\rho_0}{dE}$$
 so that 

\begin{equation}
\label{tdres1}
T^{dres} \equiv T + \Delta T = \left( \frac {d \ln \rho}{ d E } \right)^{-1} = \frac{\sigma_E^2}{E_c - E}
\end{equation}
and finally, from Eq.~(\ref{sig0})

\begin{equation}
\label{tdres}
\frac{\Delta T}{T}  = \frac{\widebar{(\Delta E)^2}}{\sigma_0^2}.
\end{equation}

As one can see, the relative shift of temperature turns out to be independent of the eigenenergy $E^\alpha$ and dependent only on the ratio between the variance $\sigma_0^2$ of the unperturbed DOS and the 
average variance $\widebar{(\Delta E)^2}$ of the SF. Again, both can be found without the diagonalization of $H$. 
The analytical values of the temperature shift for the eigenstates in Fig.~\ref{fmi04} agree fairly well with those obtained with the use of the energy shifts from Eq.~(\ref{she}) when the eigenenergy is far from the edges of the spectrum (top panels in Fig.~\ref{fmi04}).
All numerical values of the shifts are reported for the reader convenience in Table I.

\begin{widetext}
\onecolumngrid
\begin {table}[t]
\label{tabb}
\caption {Parameters for Fig.~2 of the main text. 
$E^\alpha$ is the eigenenergy, $\beta, z$ are the corresponding parameters of the BED with
$E=E^\alpha$. $E^{dres}$ is the corresponding dressed energy computed from the numerical value
of $\Delta_\alpha$. $\beta^{dres}, z^{dres}$ the corresponding 
parameters of the BED with
$E=E^{dres}$. $\Delta T/T$   is the relative temperature shift obtained from the dressed values,
$\widebar{(\Delta E)^2}/\sigma_0^2$ is the analytical value obtained from Gaussian approximation.
} \label{tab:tit} 
\begin{center}
\begin{tabular}{|c|c|c|c|c|c|c|c|c|c|c|c|}
\hline
$V=0.1$ & $E^\alpha$ & $\beta=1/T$  & $T$ & $z$ & $E^{dres}=E^\alpha +\Delta_\alpha$ & $\beta^{dres}=1/T^{dres}$ & $T^{dres}$ & $z^{dres} $
& $\Delta T/T$ &$\widebar{(\Delta E)^2}/\sigma_0^2$\\
\hline
a)   & 28.51 & 0.055 & 18.18 & 0.471 & 28.87 & 0.0525 & 19.05 & 0.464 & 0.048 &0.031\\
\hline
b)   & 11.27 & 0.2365 & 4.23 & 0.724 & 12.38 & 0.215 & 4.65 & 0.704 & 0.099 & -\\
\hline
$V=0.4$ & & & & & & & & & &\\
\hline
c) & 25.93 & 0.0735 & 13.60  & 0.505  & 30.504  & 0.0417 & 23.98  &  0.446 & 0.763 & 0.51\\
\hline
d) & 6.05 & 0.434 &  2.30 &0.873  &  16.59 & 0.156 &  6.41 & 0.637  & 1.787 & - \\
\hline
\end{tabular}
\end{center}
\end {table}

\end{widetext}

Let us remark  that even when the assumption of a Gaussian form of the DOS is not correct (for example, close to the edges of the energy spectrum) the 
BED obtained with the dressed energy in Eq.~(\ref{she}) works very well as clearly shown by a 
comparison between  full curves and symbols in the bottom panels of Fig.~\ref{fmi04}).


\section{Fluctuations in the Bose-Einstein distribution} 

Above we have shown that, on average, the numerical data for $n_s$ are in  good correspondence with the BED using
a  suitable dressed energy. However, in order to claim that statistical mechanics works on the level of individual states one has also to check whether  fluctuations are statistically acceptable.
Fluctuations can emerge by varying the eigenstates in a small energy window, or by different realizations of the disordered
inter-particle potential. We have checked that the distributions of these fluctuations are similar and can be
be considered statistically equivalent.

A study of fluctuations around average values is fundamental. Indeed,
looking at the   error bars in Fig.~\ref{fmi04}b) it is clear that  they can be  very large and one
can wonder whether they can be considered statistically acceptable. Large fluctuations typically occur 
 for eigenstates with  energies close to the edges of the spectrum or for very weak inter-particle
interaction. 

\begin{widetext}

\begin{figure}[t]
\centering
\subfloat[]{\includegraphics[width = 7cm]{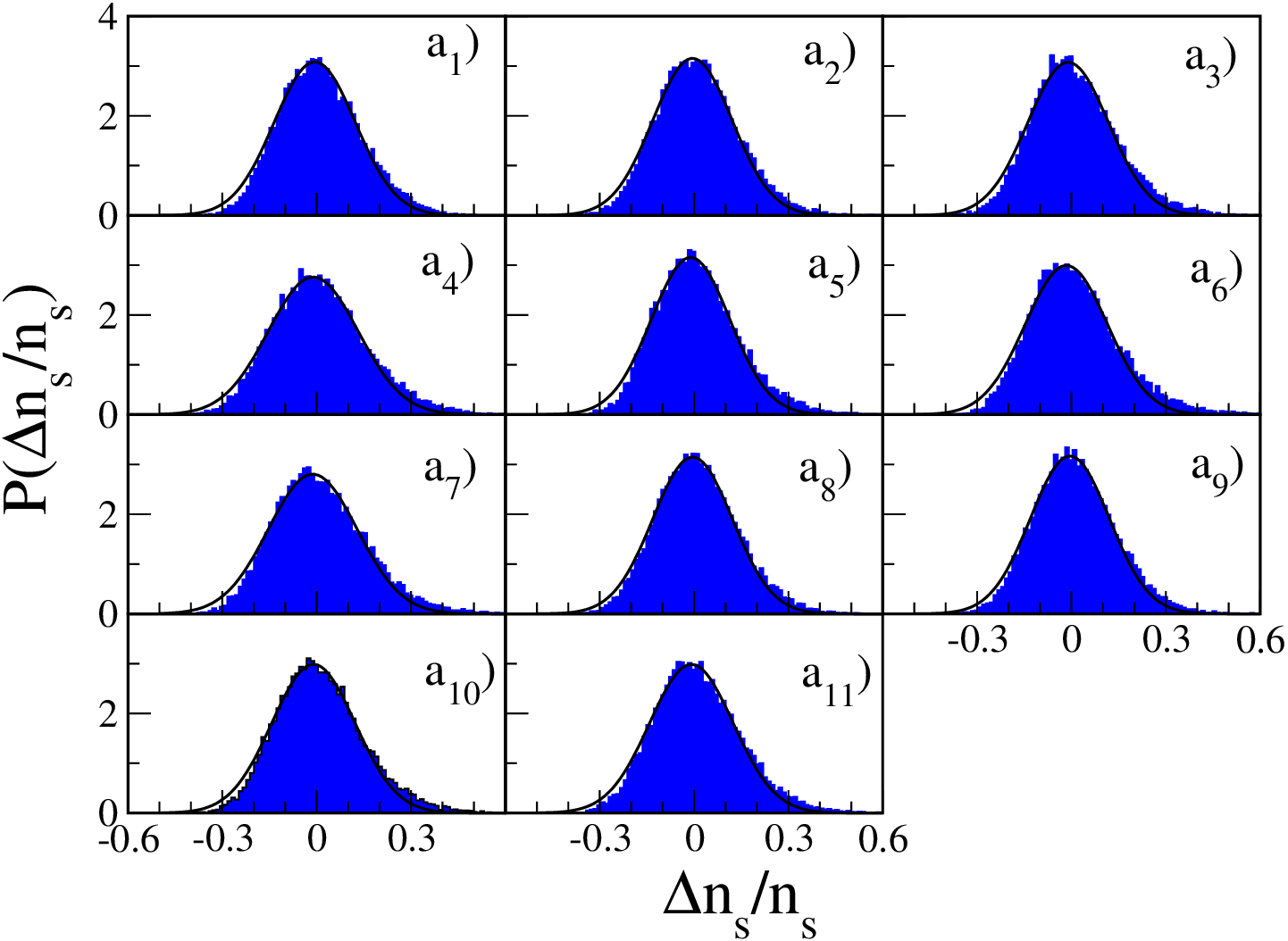}}
\hspace{1cm} 
\subfloat[]{\includegraphics[width = 7cm]{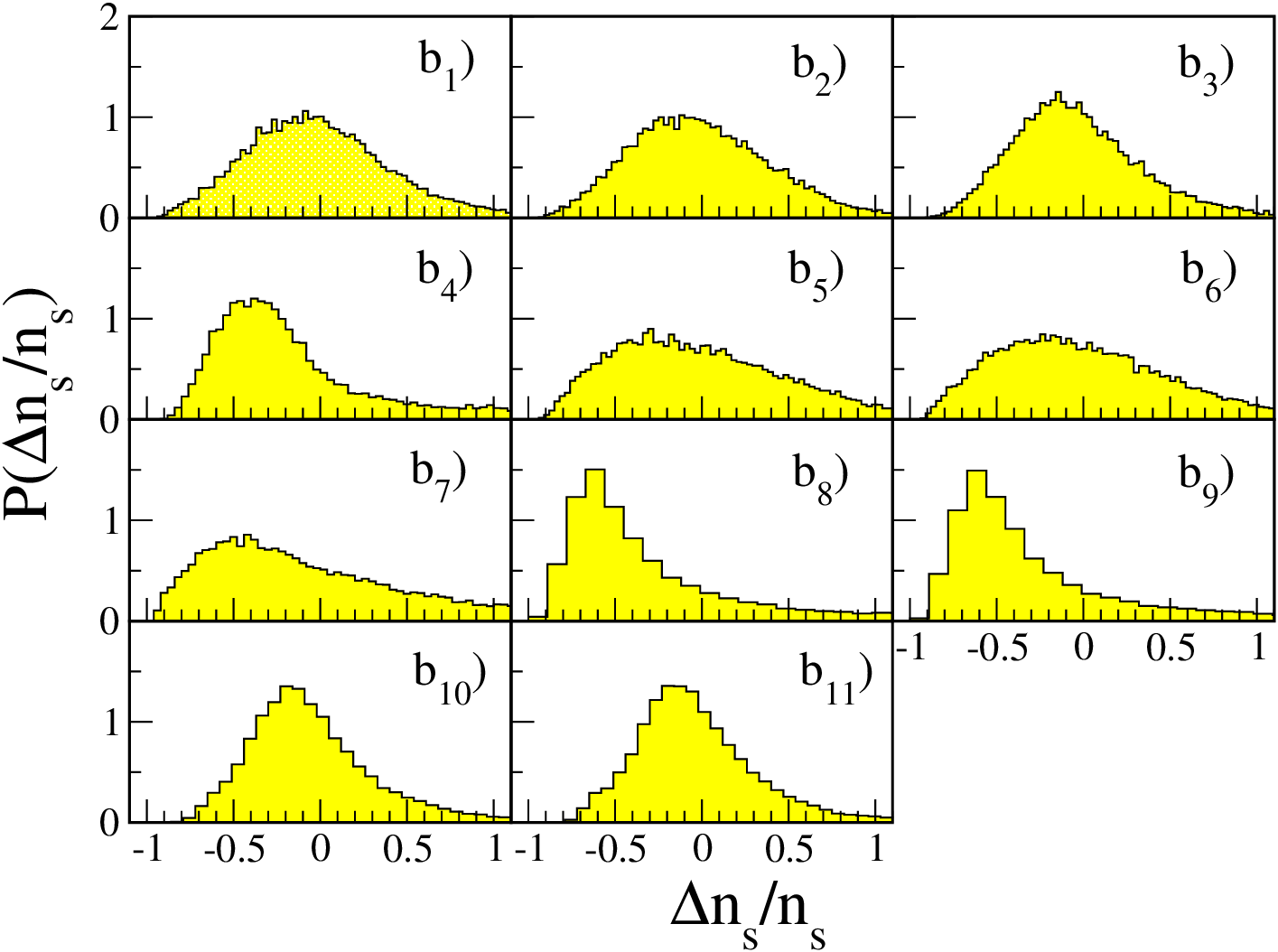}}\\
\vspace{0.5cm}
\subfloat[]{\includegraphics[width = 7cm]{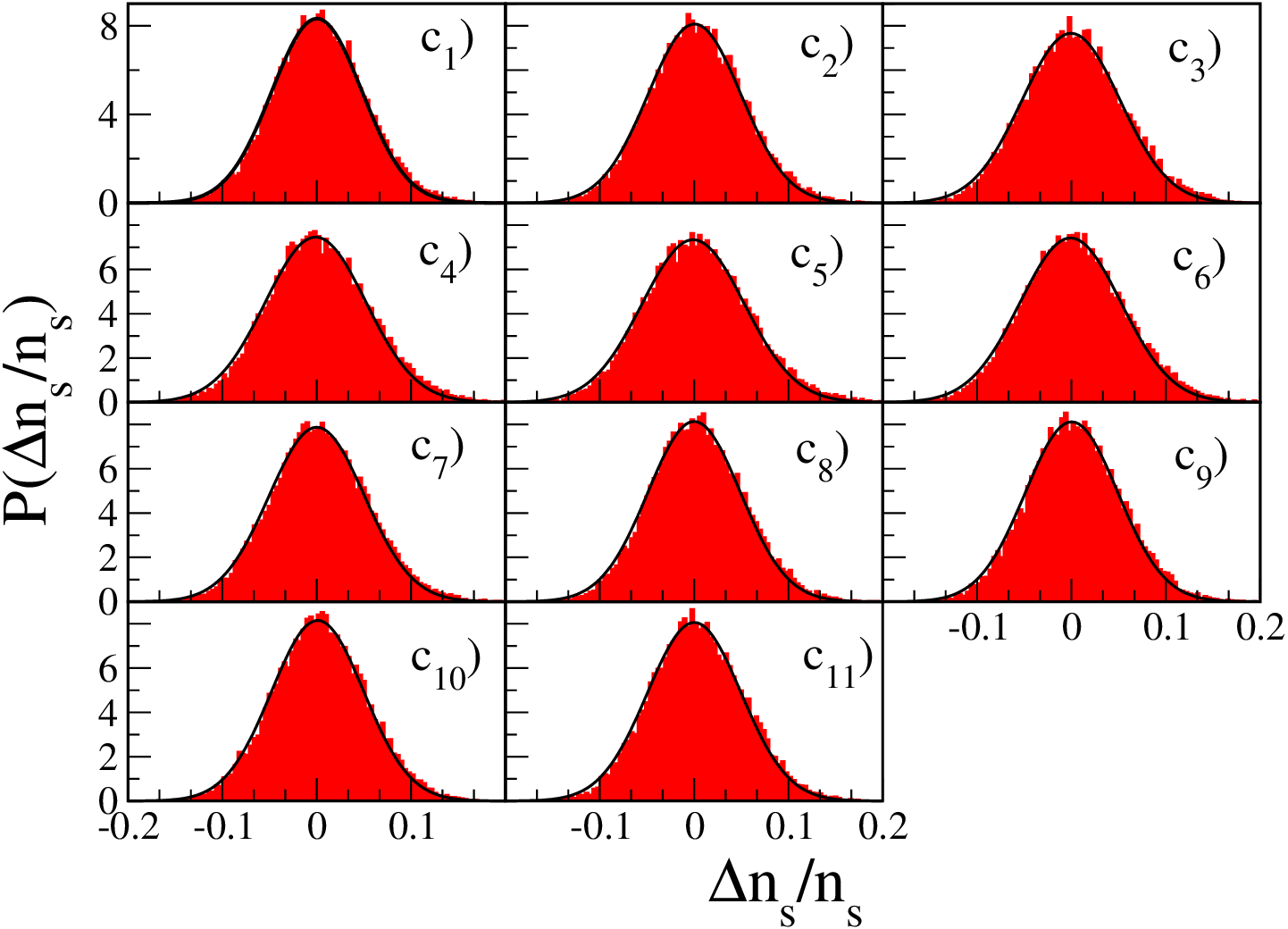}}
\hspace{1cm}
\subfloat[]{\includegraphics[width = 7cm]{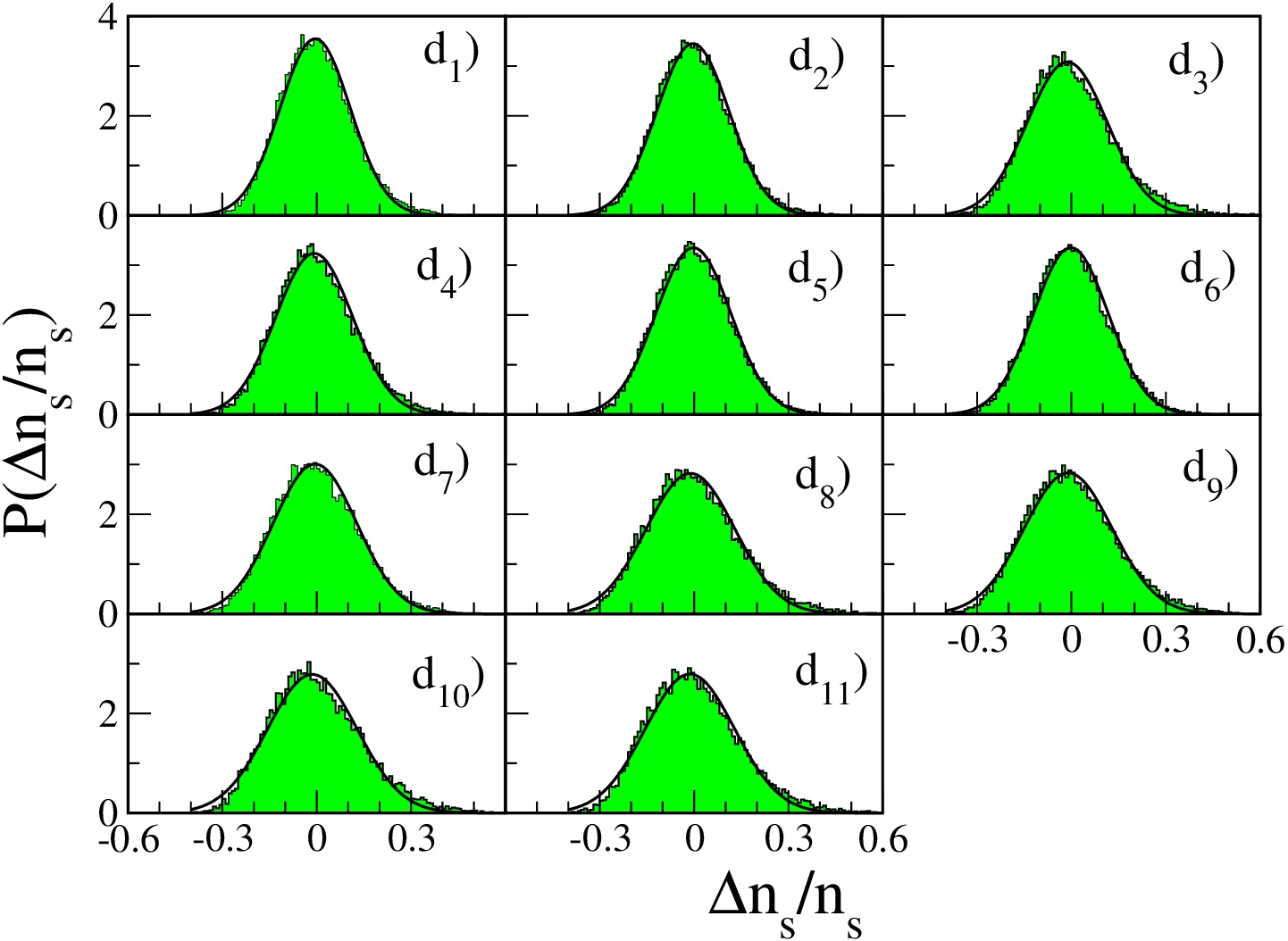}} 
\caption{(Color online)  Distributions of relative fluctuations $\Delta n_s/n_s$,
$s=1,...,M=11$:
in Fig.~\ref{fmi04}a, weak perturbation $V=0.1$, high energy,  left top panels (blue);
 in Fig.~\ref{fmi04}b, weak perturbation $V=0.1$, low energy,   right top panels (yellow);
in Fig.~\ref{fmi04}c, strong perturbation $V=0.4$, high energy,   left bottom panels (red);
in Fig.~\ref{fmi04}d, strong perturbation $V=0.4$, low energy,   right bottom panels (green).
Statistics have been obtained by 
$10^3$ different realizations of the random potential 
and choosing different eigenstates in a small energy window
in order to have approximately 20 eigenstates for each realization
of the  random potential. In all of them, where it is significant (all but the yellow distributions)
we superimposed a Gaussian fit (black curve) 
}
\label{fmi05a}
\end{figure}
 
\end{widetext}

To face the question of how relevant the fluctuations are, we studied the distribution of the relative fluctuations
$$
\frac{\Delta n_s}{n_s} \equiv \frac{n_s - \langle n_s \rangle}{\langle n_s \rangle}
$$
of the occupation numbers for close (in energy) eigenstates. 
Results are shown in Fig.~\ref{fmi05a} for different $s$-values, $s=1,..,M$ and for the four  eigenstates
of Fig.~\ref{fmi04}.

These data  clearly indicate that for all eigenstates but those in panel b), whose distributions are in the top right panels
of Fig.~\ref{fmi05a}, we have that:
\begin{itemize}
\item fluctuations  are  independent of $s$ and  therefore statistically independent,
\item  they are approximately described by Gaussians which is a strong result, in view of the requirement of standard statistical mechanics.
\end{itemize}

Concerning the eigenstates used in panel b) one can observe that for them  one has $d_{loc} \approx 0.2 > V = 0.1$.
 Therefore, applying our local
 criterion for  thermal chaotic eigenstates discussed above, we cannot treat them as chaotic eigenstates (in all other cases of course $V > d_{loc}$).

\begin{figure}[t]
\includegraphics[width=\columnwidth]{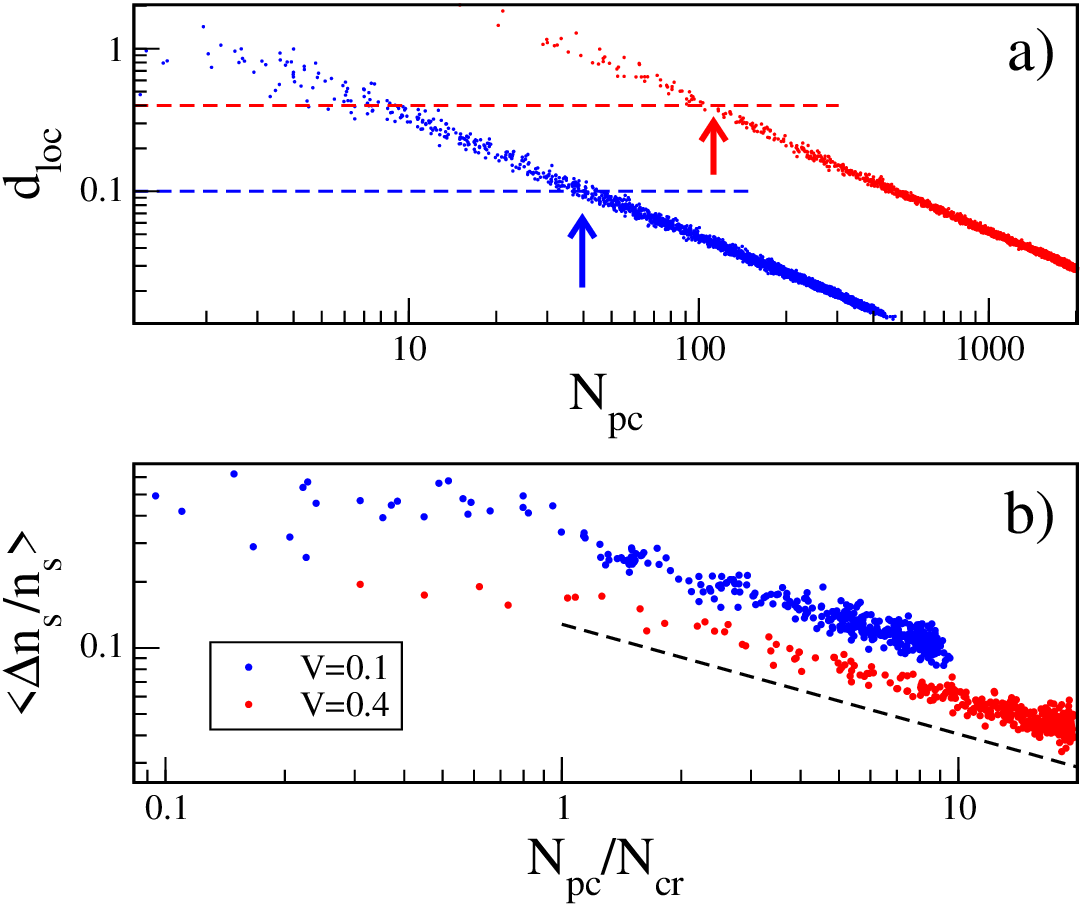}
\caption{(Color online) Top:  $d_{loc}$ as a function of the number of principal components $N_{pc}$ for each eigenstate $\ket{\alpha}$ and two different values of $V$: 0.1 (lower blue points), 0.4 (upper red points). Dashed horizontal lines represent $d_{loc} = V$. Arrows define the critical values $N_{cr}$. Bottom : average relative fluctuations in OND, $\langle \Delta n_s/n_s\rangle $  {\it vs} $N_{pc}/N_{cr}$ for two different $V$ values as indicated in the legend. Dashed line is drawn to guide the eye and stands for for $2/\sqrt{N_{pc}}$.
Average fluctuations in OND have been obtained by averaging over 20 close eigenenergies.
}
\label{fff3}
\end{figure}

For a more quantitative analysis, for each eigenstate  we have computed the corresponding value of $d_{loc}$ as a function of the number of its principal components $N_{pc}$, for the two values of $V$ considered in  Fig.~\ref{fmi04}. The intersections 
of these points 
with the horizontal lines $d_{loc}=V$, shown in Fig.~\ref{fff3}a)
 define the critical values $N_{cr}$ indicated by arrows. Then, we expect 
the fluctuations in the OND  to be not statistically acceptable when $N_{pc} < N_{cr}(V)$.  

To test such a conjecture we   group the eigenstates in small energy windows and calculating in each of them the average fluctuations in OND, $\langle \Delta n_s/n_s \rangle$. In Fig.~\ref{fff3}b) we plot such a quantity {\it vs} the renormalized number of principal components $N_{pc}/N_{cr}$. As one can see,  for $N_{pc} < N_{cr}$, the average fluctuations  $\langle\Delta n_s/  n_s\rangle$ are
almost independent of $N_{pc}$, while for $N_{pc} > N_{cr} $ they decay as $1/\sqrt{N_{pc}}$ (dashed line in Fig.~\ref{fff3}b ). This gives  a strong numerical evidence that for small systems and far from the thermodynamic limit, the value of $N_{pc}$ plays the role of an
``effective'' number of particles. 

\section {Conclusions} 

We have shown that the standard Bose-Einstein distribution, known to appear for an ideal gas in the thermodynamic limit, emerges  on the level of an individual eigenstate in an isolated system with a finite number of particles. This happens when the inter-particle interaction is strong enough to lead to the onset of chaotic many-body eigenstates in the basis defined by the chosen single-particle spectrum. In our approach we gave an analytical threshold
dependent on the eigenstates energy in order to have chaotic eigenstates. For those ``thermal eigenstates'' we computed the correspondent occupation number distribution and verified that they can be successfully described by a Bose-Einstein distribution with a suitable ``dressed`` energy dependent on the inter-particle interaction. We also gave an analytical estimate for the dressed energy  
well confirmed by direct numerical data. 

Special attention has been paid to the fluctuations of  occupation numbers. We stress that in order to have the correspondence 
with the conventional statistical mechanics, fluctuations should be  small, independent and  Gaussian. Specifically, they decrease as the square root of the number of principal components in chaotic eigenstates. Therefore, for finite isolated systems that are far from the thermodynamic limit, the control parameter for the onset of thermalization is not the number of particles but the number $N_{pc}$ of principal components in chaotic eigenstates. Our analytical findings are complemented by numerical data.

\section*{Acknowledgements} 

 This work was supported by the VIEP-BUAP grant IZF-EXC16-G. The Authors acknowledge useful discussion with L.F.Santos and R. Trasarti-Battistoni.

\appendix

\section{Occupation number distribution for a single eigenstate}

\begin{figure}[t]
\includegraphics[width=7cm]{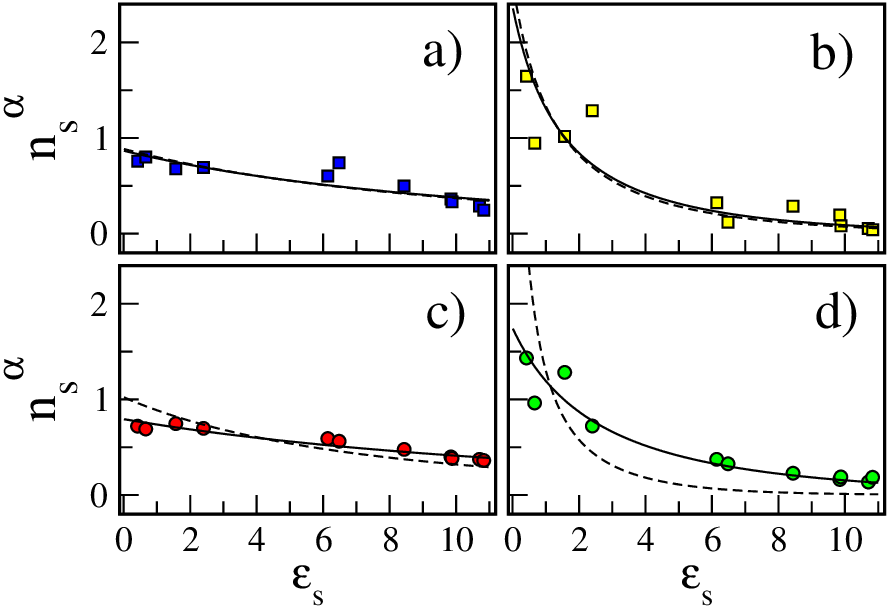}
\caption{(Color online) Occupation numbers for a single eigenstate $ n_s^\alpha $ for weak ($V=0.1$, top panels) and strong ($V=0.4$, bottom panels) perturbation and different energies:  middle of  the energy spectrum  : a) $E^\alpha = 28.51$,  c) $E^\alpha = 25.93$; edges : b) $E^\alpha = 11.27$ and  d) $E^\alpha = 6.05$.   Dashed curves are obtained from the BED with $E=E^\alpha$ in Eq.~(\ref{fco}). Full curves correspond to the BED with the energy $E^{dres}$ in Eq.~(\ref{she}).}
\label{fmi04single}
\end{figure}

It is important to observe that the occupation number distribution can be obtained
for a single eigenstate, as claimed in the title. The use of averaging over
disorder or over close eigenstates, made in the main text, had the only purpose
to introduce and analyze statistical errors.
Examples of OND for four different eigenstates, for the same parameters and energy regions
of those in Fig.~\ref{fmi04}, are shown in Fig.~\ref{fmi04single}.

\section{Properties of F-functions and density of states}

Let us start with the conventions used in the definitions of
unperturbed ($H_0$) and perturbed $(H=H_0+V)$ Hamiltonian
\begin{equation}
\begin{array}{lll}
 H_0 |k\rangle &= E^0_{k} |k\rangle \\
&\\
 H |\alpha \rangle &= E^\alpha |\alpha \rangle,
\end{array}
\label{def}
\end{equation}
and the change of representation,
\begin{equation}
  \ket{\alpha} = \sum_k C_k^{\alpha} \ket{k}.  
\label{def1}
\end{equation}
The F-function, and the strength function , defined as 
\begin{equation}
\label{ld0}
F^\alpha (E) = \sum_k |C_k^\alpha |^2 \delta(E-E^0_k),
\end{equation}
and
\begin{equation}
F_k(E) = \sum_{\alpha} |C^{\alpha}_k|^2 \delta (E-E^{\alpha}),
\label{ld11}
\end{equation} 
satisfy many relations, well known in literature (see 
for instance Ref.\cite{our2016}) and reported here
for the reader convenience. First of all they are
both normalized, 
\begin{equation}
\label{norm}
\int \ dE \ F^\alpha (E) = \int \ dE \ F_k (E) = 1.
\end{equation}
Introducing the total density of states (DOS):
\begin{equation}
\label{dos}
\rho(E) = \sum_\alpha \delta(E-E^\alpha) 
\end{equation}
and the unperturbed one
\begin{equation}
\label{dos0}
\rho_0(E) = \sum_k \delta(E-E^0_k), 
\end{equation}
both normalized to the dimension of the Hilbert space $N_H$
\begin{equation}
\label{normdos}
\int \ dE \ \rho_0(E) = \int \ dE \ \rho_0(E) =  N_H, 
\end{equation}
it is possible to write : 
\begin{equation}
\label{ld00}
F^\alpha (E) \simeq \rho_0 (E) \langle |C_k^\alpha |^2 \rangle_{E^0_k=E}
\end{equation}
where the average is performed over a number of unperturbed eigenstates with energy close to $E$.
In the same way, we have 
\begin{equation}
\label{ld01}
F_k (E) \simeq \rho (E) \langle |C_k^\alpha |^2 \rangle_{E^\alpha=E}
\end{equation}
where the average is performed over a number of eigenstates with energy close to $E$.
Note that instead of averaging over a number of eigenstates, one can use the average
inside an individual eigenstate with the "window moving" method, provided
this eigenstate has many uncorrelated components.

The two-body random interaction potential $V$ is assumed to be non-effective 
on the diagonal, i.e.
$$
\langle k | V | k \rangle =0.
$$
From these simple relations we can get 
various results concerning 
the moments of the distributions.

\subsection{First moment of the SF}
The following equalities holds,
\begin{equation}
\begin{array}{lll}
\langle E \rangle_k &= \sum_\alpha E^\alpha | C_k^{\alpha}|^2 = 
\sum_\alpha E^\alpha \langle k | \alpha \rangle \langle \alpha | k \rangle & \\
&\\
&=  \langle k | H | k \rangle = H_{kk} = \langle k | H_0 | k \rangle = E^0_{k}\\
\label{1msf}
\end{array}
\end{equation}

\subsection{Second moment of the SF}
\begin{equation}
\begin{array}{lll}
 (\Delta E_k)^2 &= \sum_\alpha (E^\alpha - \langle E \rangle_k )^2 | C_k^{\alpha}|^2 \\
&\\
&= 
\sum_\alpha (E^\alpha)^2  \langle k | \alpha \rangle \langle \alpha | k \rangle - 
\langle E \rangle_k^2 \\
&\\
&= \langle k | H^2| k \rangle - (H_{kk})^2 \\
&\\
 &= \sum_j \langle k | H | j \rangle \langle j | H | k \rangle  - (H_{kk})^2  \\
&\\
&= \sum_{j\ne k}   H_{kj}^2
\label{2msf}
\end{array}
\end{equation}

\subsection{First moment (center)  of the perturbed and unperturbed spectrum (DOS) }
Let us define $E_c$ the center of the perturbed spectrum and $E_c^0$ the center
of the unperturbed one. It is easy to see that in
our model they coincide. Indeed,
\begin{equation}
\begin{array}{lll}
 \widebar{E} &= E_c = \frac{1}{N_H} \sum_\alpha E^\alpha =  \frac{1}{N_H} Tr [H] =  \frac{1}{N_H} \sum_k 
\langle k | H | k \rangle \\
&\\
&= \frac{1}{N_H} \sum_k 
\langle k | H_0 | k \rangle = \frac{1}{N_H} \sum_k E^0_k = E_c^0\\
\label{cos}
\end{array}
\end{equation}

\subsection{Second moment   of the DOS }
\begin{equation}
\begin{array}{lll}
 \widebar{E^2} &= \frac{1}{N_H} \sum_\alpha E_\alpha^2 =  \frac{1}{N_H} Tr [H^2] 
=  \frac{1}{N_H} \sum_k 
\langle k | H^2 | k \rangle \\
&\\
&= \frac{1}{N_H} \sum_k \sum_j
\langle k | H | j \rangle\langle j | H | k \rangle \\
&\\
&= \frac{1}{N_H} \sum_k H_{kk}^2  +
\frac{1}{N_H} \sum_k \sum_{j\ne k}
 H_{kj}^2
\label{2cos}
\end{array}
\end{equation}

\subsection{Relation between second moments   of the DOS }
Let us define $\sigma_E^2$ and $\sigma_0^2$ the variances of the perturbed and unperturbed
DOS. Then, the following relation hold,
\begin{equation}
\begin{array}{lll}
 \sigma_E^2  &=  \widebar{E^2} - \widebar{E}^2 = 
  \frac{1}{N_H} \sum_k H_{kk}^2  +
\frac{1}{N_H} \sum_k \sum_{j\ne k}  
 H_{kj}^2 \\
&\\
&- (E_c^0)^2  =  \frac{1}{N_H} \sum_k (E^0_k)^2 -
\left( \frac{1}{N_H} \sum_k  E^0_k \right)^2  + \\
&\\
&\frac{1}{N_H} \sum_{j\ne k} (\Delta E_k)^2 \equiv  \sigma_0^2  + 
\widebar{ (\Delta E)^2 } \label{2rel}
\end{array}
\end{equation}
where in the latter equality we have defined $\widebar{ (\Delta E)^2 } $ as the 
average of the variances of all SF (each SF is defined for 
any given basis state $|k\rangle$).

\end{document}